\documentclass[letterpaper,english,reprint, aps]{revtex4-1}
\usepackage[T1]{fontenc}
\usepackage[latin9]{inputenc}
\usepackage{babel}
\usepackage{amsmath}
\usepackage{amssymb}
\usepackage[unicode=true]
 {hyperref}

\makeatletter

\providecommand{\tabularnewline}{\\}

\makeatother

\begin{document}
\title{Random walker derivation of Archie's law}
\author{Clinton DeW.\,Van Siclen}
\email{cvansiclen@gmail.com}

\address{1435 W 8750 N, Tetonia, Idaho 83452, USA}
\date{9 February 2025}
\begin{abstract}
Theoretical justification is provided for Archie's law. This phenomenological
equation, having the form of a power law, relates the measured electrical
resistivity of electrolyte-saturated rock samples to their connected
porosity. Historically it has been important for oil and gas exploration.
\end{abstract}
\maketitle

\section{Introduction}

The empirical Archie's law \citep{Archie} came out of the laboratory,
where electrical resistivity $\rho$ and porosity $\phi$ values were
obtained for samples of a particular rock type fully saturated by
brine or other electrolyte. The pairs of measured values $\phi$ and
$\rho/\rho_{e}$ ($\rho_{e}$ is the resistivity of the electrolyte)
were plotted on a ln\textendash ln plot and fit to a straight line,
with slope $-m$.

Thus Archie's law has the form
\begin{equation}
F=a\:\phi^{-m}\label{eq:1}
\end{equation}
where $F$ is the so-called resistivity formation factor,
\begin{equation}
F=\frac{\rho}{\rho_{e}}>1.\label{eq:2}
\end{equation}

Like the exponent $m$, the prefactor $a$ characterizes a collection
of rock samples that exhibits Archie's law. Thus Eq.\,(\ref{eq:1})
shows that the value $F\rightarrow a$ as the sample porosities $\phi\rightarrow1$.
{[}That limit is never reached since a rock sample with $\phi=1$
is nonsensical.{]} Then Eq.\,(\ref{eq:2}) reveals that the prefactor
$a>1$.

The utility of Archie's law follows from the circumstance that the
points fall on, or very near, that best-fit line. In that case Archie's
law can be used in conjunction with resistivity logs (measurements)
taken in the field, to obtain an estimate of the (conductive) porosity
of the rock formation in question (the rock matrix itself is not conductive).
An anomalously high measured resistivity value may indicate the presence
of non-conductive oil or gas in the pore space.

Here it is more convenient to consider the relation
\begin{equation}
\frac{\sigma}{\sigma_{e}}\propto\phi\,^{\mu}\label{eq:3}
\end{equation}
where $\sigma$ and $\sigma_{e}$ are the electrical conductivities
of a saturated rock sample and the electrolyte, respectively, and
$\mu$ is the cementation exponent for that sample. (In contrast,
the cementation exponent $m$ is obtained \textit{graphically} from
a collection of samples of the rock type.)

The formula for the exponent $\mu$ is obtained below in a manner
that makes clear its value is determined by the particular heterogeneity
of the porous rock. The derivation relies on the walker diffusion
method (WDM), which is introduced in the following section. Section
III then presents the derivation of $\mu$, and Sec.\,IV discusses
and gives meaning to the parameters that emerge from the derivation.
Section V gives a brief discussion of the Menger and Sierpinski sponges,
which are recursive\textemdash not ``natural''\textemdash fractals,
mainly to show example sets of the parameter values. Section VI shows
that adherence to Archie's law signifies that the rock formation is
(statistically) self-similar, over the length scales represented in
the collection of samples. Section VII applies the model developed
in this paper to the artificial rock that is an electrolyte-saturated
assemblage of glass beads. Its porosity can be reduced in steps by
sintering. Final comments are made in Sec.\,VIII.

Note that, in the remainder of this paper, the symbol $\phi$ will
refer to the volume fraction composed of the \textit{conductive domain}\textemdash the
electrolyte\textemdash that spans the sample. That lets the model
apply to porous rock that is not fully saturated with electrolyte.

\section{Walker Diffusion Method}

This application of the WDM {[}\citealp{PRE99},\citealp{JPA99}{]}
utilizes the relation

\begin{equation}
\sigma=\left\langle \sigma(\mathbf{r})\right\rangle D_{w}\label{eq:4}
\end{equation}
between the effective conductivity $\sigma$ of a composite material
and the (dimensionless) diffusion coefficient $D_{w}$ obtained from
walkers diffusing through a virtual replica of the composite. The
factor $\left\langle \sigma(\mathbf{r})\right\rangle $ is the volume
average of the constituent conductivities; the vector $\mathbf{r}$
locates a point in that volume.

The phase domains that make up the composite are host to walker populations,
where the walker density of a population is proportional to the conductivity
value of its host domain. The principle of detailed balance ensures
that the population densities are maintained, by providing the following
rule for walker diffusion over the digitized composite: a walker at
site (or pixel/voxel) \textit{i} attempts a move to a randomly chosen
adjacent site \textit{j} during the time interval $\tau=\left(4d\right)^{-1}$,
where \textit{d} is the Euclidean dimension of the system; this move
is successful with probability $p_{ij}=\sigma_{j}/\left(\sigma_{i}+\sigma_{j}\right)$,
where $\sigma_{i}$ and $\sigma_{j}$ are the conductivities of sites
\textit{i} and \textit{j}, respectively. The path of the walker thus
reflects the composition and morphology of the domains that are encountered.

The diffusion coefficient $D_{w}$ is calculated by use of the equation
\begin{equation}
D_{w}=\frac{\left\langle R(t)^{2}\right\rangle }{2\,d\,t}\label{eq:5}
\end{equation}
where the set $\left\{ R\right\} $ of walker displacements, each
occurring over the time interval $t$, must have a Gaussian probability
distribution that is necessarily centered well beyond a displacement
value $\xi$. The correlation length $\xi$ is identified as the length
scale above which a composite material attains its ``effective'',
or macroscopic, value of a scalar transport property, such as electrical
conductivity.

For displacements $R<\xi$, the walker diffusion is \textit{anomalous}
rather than Gaussian, due to the heterogeneity of the composite at
length scales less than $\xi$. Importantly, $\xi$ is expressed in
units of $\xi_{0}$, which may be considered the resolution, or size
of the smallest feature, of the system. Then a walker displacement
$\xi,$ requiring a travel time $t_{\xi}=\xi^{2}/\left(2\,d\,D_{w}\right)$,
is produced by a walk comprising $\left(\xi/\xi_{0}\right)^{d_{w}}$
segments of length $\xi_{0}$, each requiring a travel time of $t_{0}=\xi_{0}^{2}/\left(2\,d\,D_{0}\right)$,
where $D_{0}$ is the (dimensionless) walker diffusion coefficient
associated with displacements $R=\xi_{0}$. Thus $t_{\xi}=\left(\xi/\xi_{0}\right)^{d_{w}}t_{0}$,
which gives the relation
\begin{equation}
D_{w}=D_{0}\left(\frac{\xi}{\xi_{0}}\right)^{2-d_{w}}\label{eq:6}
\end{equation}
between the walker diffusion coefficient $D_{w},$ the correlation
length $\xi$ (expressed in units of $\xi_{0}$), and the parameter
$d_{w}$ associated with walker paths of displacement $\xi$.

Note that a random walk through an infinite, homogenous, conducting
space produces values $d_{w}=2$ and $D_{w}=D_{0}=1$.

\section{Derivation of the exponent $\mu$}

As porous rock is heterogeneous, the linear dimension $L$ of a sample
is the correlation length $\xi$.

Then by combining Eqs.\,(\ref{eq:4}) and (\ref{eq:6}), the effective
conductivity $\sigma$ of a rock sample of linear dimension $\xi$
is shown to be
\begin{equation}
\sigma=\sigma_{e}\,\phi\thinspace D_{0}\left(\frac{\xi}{\xi^{0}}\right)^{2-d_{w}}.\label{eq:7}
\end{equation}

Let the volume of the conductive domain spanning the sample be designated
by $V$, expressed in units of $\left(\xi_{0}\right)^{3}$. Then the
volume fraction $\phi$ is
\begin{equation}
\phi=\frac{V}{\xi^{3}}=\frac{\frac{V}{\left(\xi_{0}\right)^{3}}}{\left(\frac{\xi}{\xi_{0}}\right)^{3}}.\label{eq:8}
\end{equation}
A more useful expression of this relation is
\begin{equation}
\phi=\frac{\left(\frac{\xi}{\xi_{0}}\right)^{D}}{\left(\frac{\xi}{\xi_{0}}\right)^{3}}=\left(\frac{\xi}{\xi_{0}}\right)^{D-3}\label{eq:9}
\end{equation}
with the exponent $D$ given by
\begin{equation}
D=3+\frac{\ln\phi}{\ln\left(\frac{\xi}{\xi_{0}}\right)}.\label{eq:10}
\end{equation}
Note that the value $D<3$ reflects the relationship between sample
size and porosity. Thus $D$ can be regarded as a fractal ``mass
dimension'', indicating the extent, or volume fraction, of the conductive
domain within the sample.

Then combining Eqs.\,(\ref{eq:7}) and (\ref{eq:9}) gives the power-law
relation
\begin{equation}
\frac{\sigma}{\sigma_{e}}=\phi\thinspace D_{0}\left(\frac{\xi}{\xi^{0}}\right)^{2-d_{w}}=D_{0}\,\phi\,^{\mu}\label{eq:11}
\end{equation}
with the cementation exponent $\mu$ being
\begin{equation}
\mu=1+\frac{d_{w}-2}{3-D}.\label{eq:12}
\end{equation}
The fraction in this expression is always positive, so imposing the
lower bound $\mu>1$.

\section{Understanding $\mu$, $d_{w}$ and $D_{0}$}

Evidently the conductive domain of a rock sample is parameterized
by $D$, which is directly related to the volume fraction occupied
by that domain. In contrast, the parameter $d_{w}$, which appears
in Eq.\,(\ref{eq:6}), is a measure of the resistance to current
flow through the conductive domain. A larger $d_{w}$ value indicates
greater resistance. Note that a $d_{w}$ value is not unique to a
particular domain morphology.

The values of $d_{w}$ and $D_{0}$ cannot be determined by physical
inspection. However,\textit{ comparative} values can be inferred by
considering the behavior of walkers confined to a virtual replica
of the domain.

The parameter $d_{w}$, as introduced in Sec.\,II, is determined
by random walks that begin at one end of the rock sample and end at
the opposite end. Then the value of $d_{w}$ is obtained from the
relation
\begin{equation}
\left(\frac{\xi}{\xi_{0}}\right)^{d_{w}}=\left\langle n\right\rangle \label{eq:13}
\end{equation}
where $n$ is the number of steps of size $\xi_{0}$, comprising a
walk, and the average value $\left\langle n\right\rangle $ is obtained
from many such walks.The more tortuous and complex (e.g., presence
of dead ends) the domain morphology, the more steps are taken and
so the larger the $d_{w}$ value. In any case a finite domain is characterized
by a value $d_{w}>2$.

The parameter $D_{0}$ is the (dimensionless) walker diffusion coefficient
obtained from random walks of displacement $\xi_{0}$ (such as comprise
the displacement $\xi$). Walks confined to the conductive domain
produce $D_{0}<1$, due to the attempted moves by the walkers to exit
the domain (a consequence of the ``blind ant'' rule imposed by the
principle of detailed balance). Thus the value $D_{0}$ is smaller
for more-complex domains.

There is no algebraic relation between $d_{w}$ and $D_{0}$, but
logically a smaller value for $D_{0}$ will occur with a greater value
of $d_{w}$. For infinite, homogeneous, conductive media, $d_{w}+D_{0}=3$.
This relation may persist as a reasonable expectation for heterogeneous
and fractal media, with the conditions that $d_{w}>2$ and $D_{0}<1$.

Note that the exponent $\mu>2$ when $d_{w}+D>5$, with the condition
that $D<3$.

In any case, the value of $\mu$ is determined by the \textit{ratio}
in its expression, Eq.\,(\ref{eq:12}). The numerator accounts for
the shape or morphology of the conductive domain, and the denominator
accounts for the volume fraction $\phi$ that domain occupies.

Note that Eq.\,(\ref{eq:13}) does not impose an upper limit on the
value of $d_{w}$. Thus there is no upper limit for the value $\mu$.

High values for $\mu$ or $m$ obtained experimentally can be understood
by noting that the parameter $D$ for a given sample is determined
solely by the volume of the conductive domain. That domain may have
any one of many possible configurations. Thus high $\mu$ and $m$
values indicate a more complex (high $d_{w}$ value) domain structure.

\section{Self-similar fractals}

A heuristic example is given by the Menger Sponge \citep{Menger},
which is a fractal object in 3D space often used as a model for porous
media.

It is a recursive, self-similar fractal, such that sponges produced
by different numbers of iterations can be regarded as different ``samples''.
Equation (\ref{eq:9}) is reproduced by combining Eqs.\,(6)\textendash (8)
in Ref.\,\citep{Menger}, where $v_{i}$ is the porosity $\phi$
of the \textit{i}-th iteration sponge, and the fractal dimension $\mathcal{H}$
is $D$. And Eq.\,(\ref{eq:12}) giving the exponent $\mu$ is identical
to Eq.\,(11) in Ref.\,\citep{Menger}, giving the exponent $t$.
Note that the Menger Sponge has the characteristic length $\xi_{0}=1$.

Table I shows the calculated values of the parameters and the exponent
$\mu$, for the Menger Sponge \citep{Menger} and, for comparison,
the Sierpinski Sponge \citep{Sierpinski}. Note that the prefactor
$D_{0}<1$ reflects the condition that walkers are confined at all
porosities.

\begin{table}[h]
\caption{Calculated parameters}
\begin{tabular}{|c|c|c|}
\hline 
 & Menger Sponge & Sierpinski Sponge\tabularnewline
\hline 
\hline 
$D_{0}$ & 0.65564 & 0.935312\tabularnewline
\hline 
$d_{w}$ & 2.16326 & 2.02026\tabularnewline
\hline 
$D$ & 2.72683 & 2.96565\tabularnewline
\hline 
$\mu$ & 1.59744 & 1.58976\tabularnewline
\hline 
\end{tabular}

\end{table}

An extreme example comes from percolation theory. At the percolation
threshold, there are conductive clusters of all sizes attached to
the conducting ``backbone'' that carries the electric current. The
mass dimension of that ``incipient infinite cluster'' in 3D space
is $D=2.52295$ \citep{percolation}, and the corresponding walker
path dimension is $d_{w}=3.84331$ \citep{percolation}. Putting these
values into Eq.\,(\ref{eq:12}) gives $\mu=4.86398$.

Interestingly, Eq.\,(\ref{eq:12}), combined with Eqs.\,(18) and
(20) from Ref.\,\citep{percolation}, gives the relation $\mu=t/\beta$,
where $t$ and $\beta$ are the critical exponents in the \textit{asymptotic}
expressions for the conductivity $\sigma$, and the fraction of sites
comprising the percolating cluster, respectively.

\section{Archie formations}

A geological formation is self-similar if samples of different size
$\xi$ have $d_{w}$ and $D$ values in common. Then they have a $\mu$
value in common, and so satisfy Archie's law {[}in the form of Eq.\,(\ref{eq:1}){]}
with cementation exponent $m=\mu$, and prefactor $a=1/D_{0}>1$,
reflecting the fact that walkers are \textit{confined} to the conductive
domains.

More likely the formation is \textit{statistically} self-similar,
meaning that the variation in $d_{w}$ and $D$ values is small.

Possibly this graphical method for detecting self-similarity could
be extended to ostensibly non-self-similar (that is, heterogeneous)
formations, to look for length scales at which the formation is self-similar.
Self-similarity may be indicated when adjacent $(\phi,F)$ points
on the ln\textendash ln plot line up on a straight line that intercepts
the $\phi$ axis at $\phi>1$.

\section{Sintering experiments}

An artificial system, mimicking an unconsolidated sandstone, is a
dense, randomly packed assemblage of glass beads saturated with an
electrolyte. A consolidated sandstone is obtained by fusing the glass
beads to a lesser or greater degree by heating the assemblage above
its softening temperature for a lesser or greater time duration. Experiments
\citep{Wong} over a porosity range $0.399>\phi>0.023$ find $\mu\approx1.5$
for $\phi>0.2$ and $\mu\approx2$ for $\phi<0.2$.

Other experiments of this sort \citep{Johnson,Guyon} have obtained
similar results; in particular, finding that $\mu$ increases from
$1.5$ as the porosity $\phi$ decreases.

The increase in $\mu$ must be due to the increased densification
of the glass-bead complex. That reduces both the linear dimension
$\xi$ of the sample and its porosity $\phi$. Consequently there
may be little change in the value $D$. However the value $d_{w}$
will certainly increase, due to the further constriction of the walker
paths. Under this scenario the value $\mu$ increases.

Note that the value $\mu=1.5$ found for the higher porosity complex
indicates that $d_{w}$ and $D$ satisfy the relation $2\,d_{w}+D=7$.
As $d_{w}>2$ and $D<3$, the value $d_{w}$ must be slightly larger
than $2$ and the value $D$ must be slightly less than $3$.

\section{Concluding remarks}

Archie's law has been derived, and shown to be a consequence of the
statistical self-similarity of particular geological formations. (Not
all heterogeneity is self-similar.) This required a random-walk approach,
as self-similarity, and heterogeneity in general, are indicated by
a value $d_{w}>2$ that is obtained from random walks over the conductive
domain.

Other electrical conductivity models have been developed using fractal,
percolation, and effective medium theories. Those are reviewed or
demonstrated in Refs.\,\citep{review-1},\citep{review-2},\citep{review-3},\citep{review-4}.
The variety of approaches is a response to the difficulty in relating
measurable pore-space properties to the conductivity value of an electrolyte-saturated
rock sample.

It is clear that the parameter $d_{w}$ is responsive to the morphology
of the conductive domain. Interestingly, the walker behavior within
the domain also determines the electrical potential field \citep{PRE02}.
At equilibrium (no field), the principle of detailed balance ensures
that walkers spend equal time at every location (site) within the
domain. Then the potential field induced by application of a potential
drop across the sample is obtained by injecting random walkers at
one end of the conductive domain and immediately removing them when
they reach the opposite end. In that case the accumulated residence
times at the sites comprising the domain give the potential field.
That is, the accumulated residence time at a location within the conductive
domain is proportional to the electrical potential at that location.

The electric current flows through the sample as directed by the potential
field. While the walkers visit \textit{every} location within the
domain, and so determine the potential at \textit{every} location,
there may be conductive regions where current does not flow. This
occurs, for example, in dead-end features, where a walker can only
exit where it enters. Thus the parameter $d_{w}$ in Eq.\,(\ref{eq:12})
accounts for the tortuosity of the electric field lines, \textit{and}
for any ``dead space'', within the conductive domain.

Therefore the value of the parameter $d_{w}$ reflects both the physical
morphology of the conductive domain, and its consequent electrical
properties.

In contrast to the parameter $D$, which in principle or practice
can be calculated by the box-counting method \citep{box-counting},
there is no means to calculate $d_{w}$ for rock samples. Note that
$d_{w}$ could be calculated for the self-similar objects considered
in Sec.\,V only because those objects are \textit{infinite in size},
so allowing use of the \textit{infinitesimally small} value $\xi_{0}=1$
in calculations.

Nevertheless, it should be appreciated how remarkable it is that just
two parameters, $d_{w}$ and $D$, are sufficient to determine the
electrical properties of very complex conductive domains within pore
spaces.

\hfill{}
\begin{acknowledgments}
I thank Professor Indrajit Charit (Department of Nuclear Engineering
\& Industrial Management) for arranging my access to the resources
of the University of Idaho Library (Moscow, Idaho).
\end{acknowledgments}

\end{document}